\documentclass[journal,final]{IEEEtran}%

\normalsize

\newcommand{\exclude}[1]{}

\usepackage{times}
\usepackage{amssymb}
\usepackage{amsfonts}
\usepackage[cmex10]{amsmath}
\usepackage{dsfont}
\usepackage{url}
\usepackage{subfigure}
\usepackage{fancyvrb}

\ifCLASSINFOpdf
\usepackage[pdftex]{graphicx}
\usepackage[pdftex,colorlinks,%naturalnames,%
           bookmarks=true,%
           pdftitle=Quality\ of\ Service\ in\ Wireless\ Cellular\ Networks\ Subject\ to\ Log-Normal\ Shadowing,%
           pdfauthor=B.\ Blaszczyszyn%
           \ -\  M.K.\ Karray,]{hyperref} 
\usepackage[numbers,sort&compress]{natbib}
\usepackage{hypernat}
\else

 \usepackage[dvips]{graphicx}
\usepackage[numbers,sort&compress]{natbib}
\fi

\graphicspath{{./}{./figures/}}

\newtheorem{Th}{Theorem}[section]
\newtheorem{prop}[Th]{Proposition}
\newenvironment{Prop}{\bf\begin{prop}\rm\em}{\end{prop}} % proposition
\newtheorem{cor}[Th]{Corollary}
\newenvironment{Cor}{\bf\begin{cor}\rm\em}{\end{cor}} % proposition
\newtheorem{res}[Th]{Result}
\newtheorem{lemma}[Th]{Lemma}
\newtheorem{fact}[Th]{Fact}
\newtheorem{exe}[Th]{Example}
\newtheorem{remark}[Th]{Remark}
\newenvironment{Remark}{\bf\begin{remark}\rm}{\end{remark}} % Fact  in italic

\newcommand{\ir}{\mathbb{R}}
\newcommand{\E}{\mathbf{E}}
\newcommand{\ind}{\mathds{1}}
\newcommand{\bbT}{\mathbb{T}}

\begin{document}

\title{%
Quality of Service in Wireless Cellular Networks Subject to 
Log-Normal Shadowing
}
\author{Bart{\l }omiej~B{\l }aszczyszyn  
and~Mohamed Kadhem Karray,~\IEEEmembership{Member,~IEEE,}%
\thanks{B. B{\l}aszczyszyn is with Inria-ENS, 
23 Avenue d'Italie, 75214 Paris, France; email: Bartek.Blaszczyszyn@ens.fr}
\thanks{M.~K.~Karray is with  Orange Labs, 38/40 rue G\'{e}n\'{e}ral Leclerc,
 92794  Issy-les-Moulineaux, France; email: mohamed.karray@orange.com}
\thanks{This paper reports the results of the research undertaken  
under 2010 contract number CRE~46146063-A012 between Inria and France
T\'el\'ecom.
Partial results were presented at IFIP WMNC'2010~\cite{shadow2010}.}}

%\markboth{IEEE Transactions on Communications}%
%{Submitted paper}

\maketitle

\vspace{-12ex}
\begin{abstract}
Shadowing is believed to degrade the quality of service (QoS) in wireless cellular networks. Assuming log-normal shadowing, and studying mobile's path-loss with respect to the serving base station (BS) and the corresponding interference factor (the ratio of the sum of the path-gains form interfering BS's to the path-gain from the serving BS), which are two key ingredients of the analysis and design of the cellular networks, we discovered a more subtle reality.  We observe, as commonly expected, that a strong variance of the shadowing increases the mean path-loss with respect to the serving BS, which in consequence, may compromise QoS.  However, in some cases, an increase of the variance of the shadowing can significantly reduce the mean interference factor and, in consequence, improve some QoS metrics in interference limited systems, provided the handover policy selects the BS with the smallest path loss as the serving one.  We exemplify this phenomenon, similar to stochastic resonance and related to the ``single big jump principle'' of the heavy-tailed log-nornal distribution, studying the blocking probability in regular, hexagonal networks in a semi-analytic manner, using a spatial version of the Erlang's loss formula combined with Kaufman-Roberts algorithm.  More detailed probabilistic analysis explains that increasing variance of the log-normal shadowing amplifies the ratio between the strongest signal and all other signals thus reducing the interference.  The above observations might shed new light, in particular on the design of indoor communication scenarios.
\end{abstract}


\section{Introduction}

Modeling of the attenuation of an electromagnetic wave 
as it propagates in space
is a major component in the analysis and design of wireless 
systems. This phenomenon, also called propagation loss,
is caused by the 
decay of the signal power with the distance from the emitter
(existing even in the
free space propagation models) 
and due to  various obstacles between emitters and 
receivers (trees, buildings, hills, etc.) present in real network
profiles. Complexity and  haphazard character of actual network
profiles makes pertinent the statistical modeling of the  propagation loss.
In this approach, the propagation loss  between an emitter and a
receiver, called  {\em path-loss},
is typically modeled by the product of the {\em distance-loss function} ---
a deterministic function of the
distance between the two antennas, which represents average
path-loss on the given distance in the network, 
and a random variable, called {\em shadowing}, that takes into account
in a statistical manner the deviation from this average, observed for
each particular pair of  emitter and receiver.
We call this model {\em path-loss with shadowing}.
The distance-loss function is commonly  assumed to be some
power of the distance, with the exponent called {\em path-loss
exponent}. The random variable of the shadowing is often assumed
to have {\em log-normal distribution}, normalized to have mean one and
parametrized by its variance or standard deviation.

Various  QoS metrics in cellular networks, 
as {\em blocking probability} for constant bit-rate (CBR)
connections and {\em spectral efficiency} for variable bit-rate (VBR) 
connections, depend on the strength (i.e., variance) of the shadowing.
It is commonly believed that an increase of the variance of shadowing 
penalizes the network performance.
The results presented in this paper shed some new light on this
problem. Namely, studying the blocking probability
(defined as the fraction of the CBR  connections that cannot be established
due to insufficient transmission resources,  
in the long run of the system) 
we have discovered that it is  not always increasing with the variance
of the shadowing. For example, 
in our model of the OFDMA hexagonal network consisting of 36 BS, 
with cell radius $0.525$km and the
path-loss exponent equal to $2.5$, the blocking probability evaluated 
at the presence of the log-normal shadowing  with
the standard deviation of 25dB 
is four times smaller than in the scenario 
with no-shadowing. Even if this spectacular example regards 
a very strong shadowing, we obtain a smaller,  but  still very significant, 
decrease of the blocking probability  for 
the shadowing with the standard deviation from 7 to 15dB, which might
be appropriate for the  indoor scenario (user-indoors,  BS-outdoors); 
cf~\cite{propag_extabst}.

In all cases,  a very strong shadowing 
ultimately makes the  blocking probability tend to~1
and this dependence indeed becomes  (as expected) monotone 
for higher path-loss exponent (larger than 4 in the considered
examples).

To explain the above, somewhat surprising, observations
and extend them to other QoS metrics
we study the impact of the  shadowing and  the path-loss exponent
on the following two key characteristics of any given mobile in the
network:
\begin{itemize}
\item  its {\em path-loss to the serving BS}, which is the 
 one received with the strongest signal (and not necessarily the
 closest one),
\item  the so called  mobile's {\em interference factor},
defined %
as the ratio of the sum of the path-gains 
form interfering BS
to the path-gain from the serving BS.
\end{itemize}
These are two key ingredients in the analysis of wireless cellular
networks and  thus their mean values can be considered 
as some QoS ``pre-metrics''.
In particular, they are explicitly present in 
the call blocking  condition --- the one 
used to control the admission of streaming users, and hence 
intrinsically related to the blocking probability.
They are, even more straightforwardly, 
related to the spectral efficiency of the data networks.
While being key ingredients  in the study of wireless systems, 
the (mean) QoS pre-metric are also more easy to analyse. In particular
they do not depend on a particular assumption regarding the
spatial correlation of shadowing
(which is not the case, e.g. for the mean blocking probability).

We have studied  the mean values of the above two basic QoS
ingredients, with the averaging taken over all possible
locations of users in the network and over the distribution of 
the shadowing.

Our main findings are as
follows:
\begin{itemize}
\item The mean path-loss (with respect to the  serving BS)
is always increasing  %
in  the variance of the shadowing. 
The ultimate degradation of the QoS for large
shadowing variance is due to this increasing path-loss.
(When possible, this may be however remedied by increasing the  power
of the emitted signals). 
\item The mean interference factor %
is  not monotonic in the variance of the shadowing.
It first increases and then decreases (asymptotically to zero!), when
the shadowing variance goes to infinity. 
This asymptotic behaviour
can be heuristically explained by the {\em single big jump
  principle} of heavy-tailed distributions:
the sum of the (log-normal) path-gains form all antennas is dominated 
by a big value of the path-gain from a single antenna. When this
antenna is the serving one, then this big path-gain does not count in the 
interference, which becomes negligible in proportion to it.
\item The above  two facts 
lead to  the phenomenon that we may call a {\em stochastic resonance for 
 QoS in path-loss-and-interference limited
systems}: when QoS is not yet compromised by  path-loss conditions, 
a moderate increase of the shadowing variance
may make it profit from the reduction of the interference.
\end{itemize}
We confirm the above findings by a mathematical analysis of the respective
stochastic models. 
We also compare in this matter
the performance of the perfect (hexagonal)
and  irregular (Poisson) networks and find that both architectures
exhibit very similar QoS ``pre-metrics'' for the standard deviation of
the shadowing  larger than 20dB. Moreover,  
we prove an interesting
invariance of the QoS metrics of the infinite
Poisson cellular networks with respect to the
distribution of the shadowing.
As a consequence we also obtain  fully explicit, analytical results
for the mean path-loss and interference factors in the case of the
infinite Poisson network.

\emph{The remaining part of this paper} is organized as follows. In
the next section  we briefly present related works. 
In Section~\ref{s.ModelDescription} we describe our models.
The main numerical results are presented
in Section~\ref{s.Numerical}. Next, in Section~\ref{s.Mathematical}
we present mathematical analysis of the models, which supports and
completes our  numerical findings. Finally, in
Section~\ref{s.Concluding} we provide some concluding remarks.

\section{Related works}
\label{s.RelatedWorks}
The propagation loss model considered in this
paper  is commonly accepted in the literature;
see e.g. \cite{Stuber2001} where log-normal shadowing of mean $1$ is
considered. 
A possible extension of this model consists in assuming shadowing
distribution (say, its variance) that depends on the distance,
cf.~\cite{Liang2008}.

The impact of the shadowing on the distribution of the interference
factor is studied numerically in~\cite{MasmoudiTabbane2004}
and analytically in~\cite{KelifCoupechoux2009Shadowing}. 
However, the above two articles do not take into account the 
modification of the network geometry induced by the shadowing, i.e.,
assume that mobiles are served by their geographically closest BS.
This is not a realistic assumption and, as we will show in this paper, 
leads to misleading conclusions that the shadowing  dramatically
increases the mean interference factor.  

The paper~\cite{ViterbiViterbiZehavi1994} 
focuses on the interference factor averaged over a given cell, 
and in particular the effect of shadowing on this average. It is
shown there that the cell shape modification induced by the shadowing affects
significantly the mean interference factor. 
More precisely, that this mean
decreases substantially if mobiles are served by the BS  offering
the smallest  path-loss. We  adopt this assumption  throughout the present
paper in the context of  regular (hexagonal) and irregular  
(Poisson)  geometry  of BS, as proposed
in~\cite{BaccelliBlaszczyszynKarray2004}.

Some papers (see
e.g.~\cite{KelifCoupechouxGodlewski2008,Karray2009FFactor})
 propose more explicit approximations of the interference factor and its moments
(mean and variance) assuming only deterministic propagation loss
models (without random shadowing).
\cite{Staehle2002} studies the distribution of the interference factor
in such a case. 

In~\cite{Minelli2010} the authors partially confirm,
by a different approach,  our 
early observation from~\cite{shadow2010}, 
that the average SIR (which is the inverse of the interference factor) 
might increase with the shadowing variance
when the best server policy is chosen.

The interference factor was recognized very early as  a key
element in the performance evaluation of cellular networks;
cf.~\cite{ViterbiViterbi1993Erlang,LeeMiller1998}. 
Fundamental to our approach to the evaluation of the blocking
probability are papers~\cite{Zander1992,Zander1992Dist}. They show
how the power allocation problem without power limitations
can be reduced to an algebraic system of  linear  inequalities.
Moreover, they recognize that the spectral radius of the (non-negative) matrix
corresponding to this system not greater than~1 is the
necessary and sufficient condition of the feasibility of power allocation
without power limitations.  
This approach lead to
the development of a comprehensive framework of the evaluation  of the
blocking probability in  CDMA, HSDPA and OFDMA,
via a spatial version of the famous Erlang's
formula in~\cite{BaccelliBlaszczyszynTournois2003,BaccelliBlaszczyszynKarray2004,BaccelliBlaszczyszynKarray2005Blocking,BlaszczyszynKarray2009OFDMA}.
QoS in data networks are studied using this approach in 
\cite{BlaszczyszynKarrayInfocom2007}.

Finally, recalling that the mean QoS pre-metrics studied in this paper
do not depend on the
spatial correlation of the shadowing, we 
remind~\cite{Gudmundson91,Klingenbrunn99} as bringing models that can
be used when studying the spatial distribution of the QoS metrics.

\section{Model description}
\label{s.ModelDescription}
\subsection{Location of base stations}
\label{ss.BSgeometry}
In this paper we will consider two particular models for the location
of BS, hexagonal and Poisson  one. The former  is
commonly considered as an  ``ideal'' model for the cellular networks,
while the latter  one can be seen as 
an extremal case of very irregular network.  

\subsubsection{Infinite Models}
\begin{itemize}
\item{\em Hexagonal network.}
\begin{figure}[h]

\centering\includegraphics[width=0.8\linewidth]{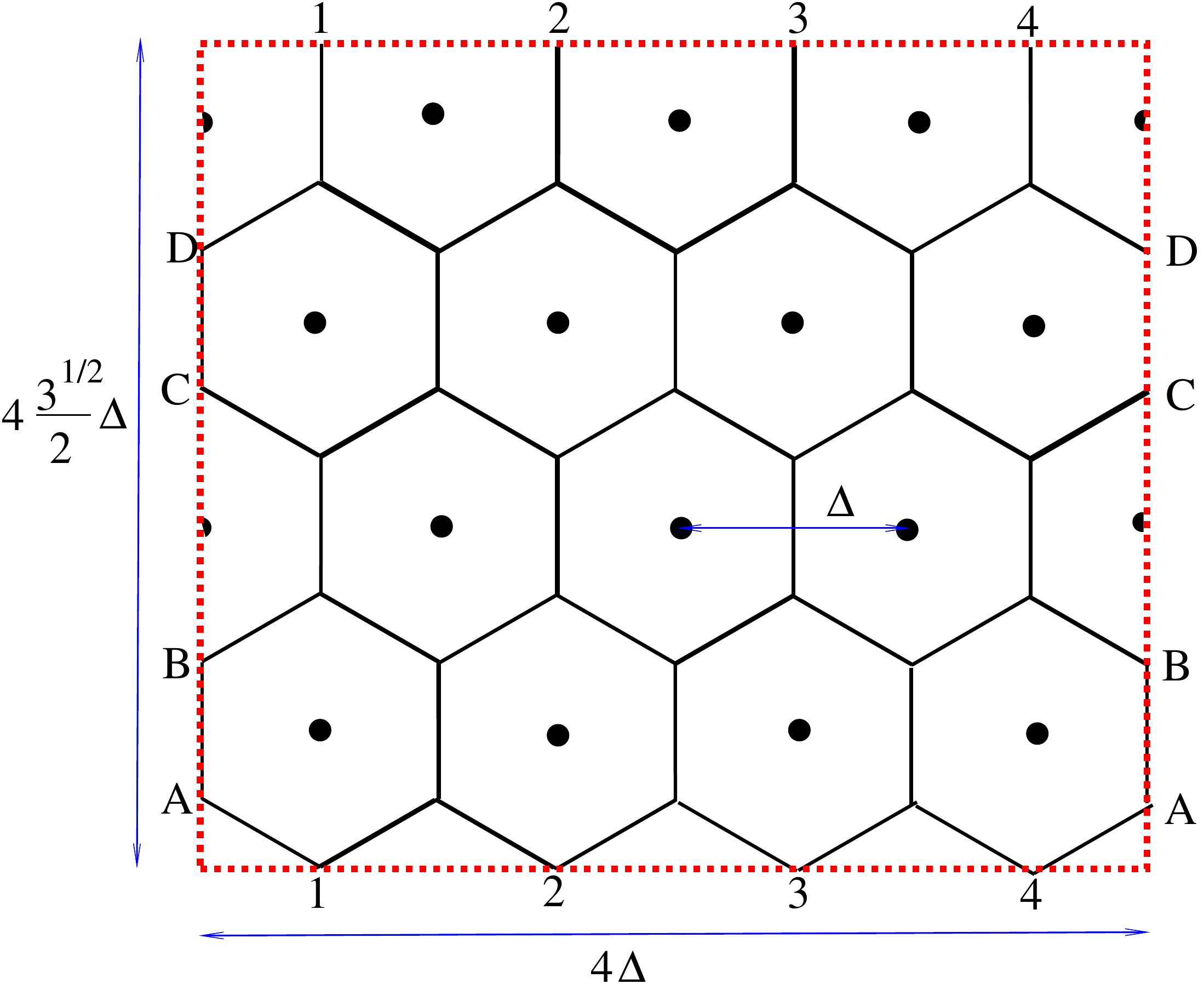}
\caption[Hexagonal pattern.]{\label{fig.T4}
Hexagonal pattern of $4\times4$ BS on rectangular torus
  $\bbT_4$. Identified points are denoted by the same digits or characters.}
\end{figure}
Consider BS located on a regular hexagonal grid on $\ir^2$
with the distance  $\Delta$ between two adjacent vertices   of this
grid \footnote{The set of vertexes of this grid can be described on the complex plane
by $\{\Delta(u_1+u_2e^{i\pi/3}),\
u=(u_1,u_2)\in\{0,\pm 1,\ldots\}^2\}$.}; cf.~Figure~\ref{fig.T4}.
Note that the surface area of a given cell (hexagon, i.e. subset of
the plane whose points are closer to a given point of the grid 
than to any other) of this model 
is equal to  $\sqrt 3\Delta^{2}/2$. Thus the intensity of the BS in
this model is equal to $\lambda=2/(\sqrt 3\Delta^2)$
BS/$\text{km}^2$.
In what follows it will be  customary to consider a stationary version $\Phi_H$
of this grid, which can be obtained 
by randomly shifting it through a vector
uniformly distributed in one given hexagon
(cf.~\cite[Example~4.2.5]{FnT1}). 
In this model a given location, say the origin of the plane,
corresponds to an ``arbitrary'' location of a mobile,  
``randomly chosen'' in the network.   
\item{\em Poisson network.} 
Assume that BS are located at the points of a stationary, 
homogeneous Poisson point process  (p.p.) $\Phi_P$
of intensity $\lambda$ BS/$\text{km}^2$ on the plane $\ir^2$.
When comparing performance of Poisson and hexagonal model we will
always take them with the same intensity $\lambda=2/(\sqrt 3\Delta^2)$.
\end{itemize}

Considering infinite models is often a convenient way of 
studying phenomena arising in very large networks. A particular 
property of these models is lack of (geographic) boundary effects, which
in real, large but finite, networks,  have often a negligible impact on
performance characteristics measured in the ``middle'' of the network.
However, as we will see in this paper, sometimes mathematical
assumption of an infinite network  may create some artifacts, which 
are not observed in more realistic, large but finite, networks. 

\subsubsection{Bounded Models}
\label{sss.BoundedModels}
In order to have finite network models, and
still neglect the boundary effects (which might be reasonable for
large networks) one often considers {\em toroidal model}.
Recall that, roughly speaking, rectangular 
torus is a rectangle  whose opposite sites are ``identified''.
For $N=2,4,6,\ldots$, we will denote by $\bbT_N$ the rectangle 
$[-N\Delta/2,N\Delta/2)\times[-N\sqrt3\Delta/4,N\sqrt3\Delta/4)$ with toroidal
metric.
 Restricting $\Phi_H$ to
$\bbT_N$, i.e. taking $\Phi_H^{\bbT_N}=\Phi_H\cap\bbT_N$ one obtains the
model whose distribution is invariant with respect to translations
on the torus. Thus we obtain a hexagonal network model that
consists of $N^2$ cells 
(cf.~Figure~\ref{fig.T4}) and which does not exhibit any border effects.
Similarly we will consider the restriction $\Phi_P^\bbT$ of the Poisson
p.p. $\Phi_P$ to $\bbT_N$.

\subsection{Path-loss model with shadowing}
\label{ss.Pathloss}
For a given BS $X\in\Phi$ ($\Phi=\Phi_P$ or $\Phi_H$) and a given
location $y\in\ir^2$ on the plane we denote by 
$L_{X}(y)$ the (time-average, i.e., averaged out over the fading) 
path-loss between BS $X$ and location $y$.
In what follows we will always assume that 
\begin{equation}\label{Lshad}
L_X(y)=\frac{L\left(  \left\vert X-y\right\vert \right)  }{S_{X}(y)}\,,
\end{equation}
where  $L(\cdot)$, called {\em distance-loss}, 
is a non-decreasing, {\em  deterministic}
function
of the distance between an emitter  and a receiver, 
and $S_{X}(\cdot)$ is a {\em random shadowing field} related to the BS $X$.
In what follows we call $L_{X}(y)$ {\em path-loss with shadowing} (or
path-loss for short) between $X$ and $y$.
Moreover, we will always assume that given locations of BS
$\{X_i\in\Phi\}$  their shadowing fields $\{S_{X_i}(\cdot)\}$ are
{\em independent} non-negative stochastic  processes, each being
indexed by locations $y\in\ir^2$. 
More formally speaking, the locations of BS $X$ and their respective
shadowing fields $S_X(\cdot)$ form an {\em independently marked}
version $\tilde\Phi=\{(X,S_X(\cdot))\}_{X\in\Phi}$ 
of the point process~$\Phi$.

Regarding the distribution of the marks (shadowing fields) of this
process, they  are assumed to have  {\em
  the same marginal distributions}, i.e.,  given $X$,   
$S_X(y)$ has the same distribution for all $y\in\ir^2$, of normalized mean  
$\E[S_X(y)]=1$, with  the following two cases  being  of particular interest
\begin{itemize}  
\item $S_X(y)\equiv1$, which corresponds to a case with negligible
  shadowing (we will say also ``no shadowing''),
\item for all $y$, $S_X(y)$ is log-nornal random variable with mean~1. 
Recall that such a mean-1 log-normal variable $S$ 
can be expressed as $S=e^{\mu+\sigma N}$ where
  $N$ is standard Gaussian random variable (with mean~0 and
  variance~1) with $\mu=-\sigma^2/2$ and some constant~$\sigma$.
 Indeed, in this case
  $\E[S]=e^{\mu+\sigma^2/2}=1$.
Note that if the shadowing is log-normal random variable
then the  path-loss (at a given distance) expressed in dB  is Gaussian
random variable.  Furthermore, in this context it is common to parametrize 
the log-normal shadowing by the standard deviation (SD) of  $S$
expressed in dB, i.e., the SD of $10\log_{10}S$. We will denote it by
$v$. With respect to the  previous parametrization we have
$v=\sigma10/\log10$. Throughout the
paper we will call $v$ the {\em logarithmic standard deviation
  (log-SD) of the
  shadowing}.
\end{itemize}
If not  otherwise specified, we do not make any particular assumption on the
correlation of the shadowing field $S_X(y)$ for given $X$ and 
different locations $y$.

Throughout the paper we will implicitly
assume also that mean {\em path-gain} is finite, i.e., $\E[1/S]<\infty$.
Note that this condition is satisfied for log-normal variable,
indeed, in our case of mean-1 variable $\E[1/S]=e^{\sigma^2}=e^{v^2\log^210/100}$. 

For the deterministic  distance-loss function $L(\cdot)$ the
following particular model  is  often used and will be our default 
assumption in this paper:
\begin{equation}\label{e.L}
L\left(r\right)  =\left(  K r\right)  ^{\beta}%
\end{equation}
where $K>0$ and $\beta>2$ are some constants.

\subsection{Handover policy and path-loss factor}
\label{ss.handover}
In what follows we will assume that each given  location 
$y\in\ir^2$ is served by the BS $X_y^*\in\Phi$ with respect to which
it has the weakest path-loss $L_{X_y^*}(y)$
(so, in other words, the strongest
received signal, given all BS emit with the same power), i.e, such that
\begin{equation}\label{e.handover}
L_{X_y^*}(y)\le  L_{X}(y)\qquad \text{for all\  }X\in\Phi\,,
\end{equation}  
with any tie-breaking rule. 
Note that in the case of negligible shadowing ($S_X(y)\equiv 1$) and 
strictly increasing function  $L(\cdot)$ the 
above policy corresponds to the geographically closest BS.
Note also that for our infinite network models with random shadowing, 
one has to prove that the minimum of the path-loss is achieved for
some BS, i.e., that $X_y^*$ is well defined. 

Note that $L_{X_y^*}(y)$  is the path-loss experienced by a user located at
$y$ with respect to its serving BS. Obviously it determines the QoS of 
this user (we will be more specific on this  in Section~\ref{ss.Blocking}).
In this context we will call it {\em path-loss factor}%
\footnote{not to be confused with the path-loss exponent $\beta$} 
of user $y$ and
denote  by   $l(y)=L_{X_y^*}(y)$. Note that it depends on the location
$y$ but also on the path-loss conditions of this location with respect
to all BS in the network $l(y)=l(y,\tilde\Phi)$.
Path-loss factor $l(y)$ is typically not enough to determine the QoS of a given user.

\subsection{Interference factor}
\label{ss.ffactor}
For a given location $y\in\ir^2$ we define the interference factor
$f(y)$ as 
\begin{equation}
\label{ffactordef}
f(y)=f(y,\tilde\Phi)=\sum_{X\in \Phi, X\not=X_y^*}
\frac{L_{X_y^*}(y)}{L_{X}(y)}=\sum_{X\in \Phi}
\frac{l(y)}{L_{X}(y)}-1
\end{equation}
provided $X_y^*$ is well defined. 

Study of the path-loss and interference factors, which are 
relatively simple objects, 
can give an important insight into more involved QoS
metrics, such as  blocking probability in streaming traffic and
mean throughput in data traffic.
In what follows we recall how
$l(y)$ and $f(y)$
appear naturally in the evaluation
of the blocking probabilities. %

\subsection{Blocking probability; a space-time scenario}
\label{ss.Blocking}
In this section we briefly describe the  
relation between the path-loss and interference factors
and the blocking probability. 
This relation, whose very essence can be
explained by the famous Erlang's loss formula, was observed
in the current geometric context (however without shadowing) 
in~\cite{BaccelliBlaszczyszynKarray2005Blocking}.

In order to evaluate the blocking probability it is
necessary to specify the dynamics of call arrivals and their durations,
as well as to identify the set of feasible configurations of users
(which  can be served simultaneously at their requested bit-rates).
To this regard, consider {\em a given  realization of
the netowrk with shadowing $\tilde\Phi$}, 
and a  spatio-temporal Poisson arrival process of 
calls which require from
 the network some predefined transmission
rates for some exponential transmission times.
These rates can be maintained at the price of
blocking of some call arrivals when a network
congestion occurs.
The fractions $b=b(\tilde\Phi)$ of blocked  arrivals in the long
run of the system is called the {\em blocking probability}.
By the famous {\em Erlang's loss formula},
it is equal to the conditional probability that in the stationary 
configuration of the (non-blocked) arrival process
the system cannot admit a new user, given all users in the current
configuration can be served.
Moreover, if the decision whether to block a given call  (or admit it) 
is based on the verification of  some {\em feasibility condition}
that has the so called {\em multi-Erlang form},
then the  Erlang's loss formula can be relatively easily  evaluated,
e.g. discretizing the values of the SINR and using
{\em Kaufman-Roberts algorithm}. 
A canonical form of the multi-Erlang feasibility condition
involves verification by each BS $X$ of the following condition 
\begin{equation}\label{eq:m-erlang}
\sum_{y: X_y^*=X}
\varphi\Bigl(l(y),f(y)\Bigr) \le 1\,,
\end{equation} 
where the summation is taken over all users (including a new arrival)  
to be served by the BS $X$ and 
$\varphi(\cdot,\cdot)$ is some function of the path-loss
and interference factors of user $y$.
This condition guarantees sufficient wireless resources
to maintain the predefined transmissions rates for all served mobiles.
Specific form of the function $\varphi(\cdot,\cdot)$
needs to be developed for each 
particular cellular technology (taking into account the performance of
the coding schemes, type of the
multiplexing, etc.). 
Below we show two  examples 
borrowed from our previous studies. They give some insight 
into how the feasibility condition~(\ref{eq:m-erlang}) 
depends on the user transmissions rates, it is 
supposed to guarantee.

\begin{itemize} 
\item For the down-link in the OFDMA network
\begin{equation}\label{eq.OFDMA-phi}
\varphi(l,f)= \frac{r}{W\psi\Bigl((1-\epsilon)/
((Nl/\tilde P)+\alpha+f)\Bigr)}\,,
\end{equation}
 where $\tilde P$ is the  maximal BS power, 
$\epsilon$ is the fraction of this maximal power used in common
  (pilot) channels, $\alpha$ is the intra-cell orthogonality factor
(usually assumed to be~0 in OFDMA), 
  $N$ external noise power, $W$ is the system bandwidth, $r$ is 
the required bit-rate $r$ of user and $\psi$ is the link performance function
($\psi(\xi)$ is the bit-rate per Hz available
when SINR is equal to  $\xi$;
\footnote{\label{fn.fading} e.g., assuming additive white Gaussian noise 
(AWGN) channel and the link performance closed to the optimal one,
  $\psi$ is  given by the famous Shannon's formula
$\psi(\xi)=\log_2(1+\xi)$. Taking $\psi(\xi)=a\log_2(1+\xi)$ with
  some constant $a\le 1$  permits to account for a
 degradation of the link performance in practical
systems compared to the ideal AWGN case; cf.~\cite{GoldsmithChua1997}.
Further extensions consider the Single-Input-Single-Output (SISO) AWGN channel 
with fading, for which  the known formula for the ergodic capacity
is $\psi(\xi)=\E[\log_2(1+\xi |F|^2)]$,  where
the expectation is with respect to the distribution of the channel
fading $F$, 
and the Multiple-Input-Multiple-Output (MIMO) AWGN channel, whose ergodic
capacity is  $\psi(\xi)=\E[\log_2\det(I+\xi \mathbf{FF}^T)]$, where
 $\mathbf{F}$ is the vector of channel fading;
 cf.~\cite{Telatar1995}.});
 cf.~\cite{BlaszczyszynKarray2009OFDMA}.
\item For the  down-link in CDMA network
\begin{equation}\label{eq.CDMA-phi}
\varphi(l,f)=\frac{\xi}{1+\alpha\xi}\frac{1}{1-\epsilon}
\Bigr(\frac{Nl}{\tilde P}+\alpha+f\Bigr)\,,
\end{equation}
where $\xi=\psi^{-1}(r/W)$ is 
the  SINR threshold  corresponding to
the required bit-rate $r$ of user and the remaining notation 
notation as above; (cf.~\cite{BaccelliBlaszczyszynKarray2005Blocking}).
\end{itemize}

In what follows we will denote by
$\E[b]=\E[b(\tilde\Phi)]$ 
the blocking probability {\em averaged over possible scenarios
regarding locations of BS and their shadowing conditions}.
It can be evaluated by the simulation of several realizations of
the network with shadowing $\tilde\Phi$, 
evaluation of  $b(\tilde\Phi)$  by
the  Kaufman-Roberts algorithm as described above, and then 
taking the empirical average over the realizations of $\tilde\Phi$.
However, in practice {\em one realization of $\tilde\Phi$ is
enough}, provided the shadowing fields $S_X(y)$ do not exhibit 
high spatial correlation across $y$ ( recall that we have assumed them to
be independent across $X$); cf. Footnote~\ref{fn.LLN}. 
Indeed, we have noticed in our experiments, that for large enough
networks (in the case of the  hexagonal network $\bbT_6$ is enough!)
with spatially uncorrelated shadowing,
the value of $b(\tilde\Phi)$ is almost  
invariant with respect to $\tilde\Phi$ and hence  very close to
$\E[b(\tilde\Phi)]$. This is due to spatial
ergodic properties of the process $\tilde\Phi$.

\subsection{Our methodology in the study of the network QoS}
\label{ss.methodologh} 
In section~\ref{ss.NResBlocking} we will show some numerical examples, 
which show the  typical
dependence of the blocking probability $\E[b]$ on the parameters of the
path-loss model. These examples, restricted to OFDMA, 
are not supposed to be exhaustive.
The goal is to show the typical tendencies.

In order to explain these tendencies, in Sections~\ref{ss.NResFFactor}
and~\ref{ss.NResLFactor} we will  study more thoroughly the
mean  values of the interference and path-loss factor 
$\E[f(y)]=\E[f(y,\tilde\Phi)]$, $\E[l(y)]=\E[l(y,\tilde\Phi)]$
(where the expectation $\E[\ldots]$ corresponds to the distribution 
of $\tilde\Phi$, i.e., this 
of the shadowing field and of the random  location of the user).
By the translation
invariance of  
the distribution of our infinite and toroidal models, these
expectations (corresponding the spatial averaging)   
do not depend on the user location and thus, for these
models, $\E[l(y)]=\E[l(0)]$ and $\E[f(y)]=\E[f(0)]$~%
\footnote{\label{fn.LLN}
Often the mathematical expectation $\E[f(0,\tilde\Phi)]$ 
(and similarly for $\E[l(0,\tilde\Phi)]$)
corresponds to  
the {\em empirical mean value} $\lim_{n\to\infty}1/n\sum 
f(y_i,\tilde\Phi)$ of 
the interference factor measured at many locations ``uniformly''
sampled in one given realization of the network and shadowing. 
A precise statement and rigorous proof of such an ergodic  result
is  beyond the scope of this paper. We remark only that  
for the hexagon network on the torus,
this result follows simply form  the Law of Large Numbers,
when $y_i$ are independently and uniformly distributed
and provided the shadowing variables $S_X(y_i)$ are independent
across different values of $y_i$. Indeed, in this case
$f(y_i,\tilde\Phi)$ and $l(y_i,\tilde\Phi)$ are independent,
identically distributed (across $i$) random variables.
However, recall that 
the latter assumption, corresponding
to spatially uncorrelated shadowing,  
is {\em not} our  default assumption,
since it is not needed for other results regarding $\E[l(0)]$ and $\E[f(0)]$.}.

Our methodological conjecture is as follows.
We believe that the {\em mean path-loss and interference factors
$\E[l(0)]$, $\E[f(0)]$
can be considered as  primitive (basic) metrics of the  QoS
and their behavior can (at least qualitatively) explain the main tendencies 
observed for more involved QoS metrics}.
This methodological conjecture is motivated 
by the observation that 
the function $\varphi$ in the feasibility condition~(\ref{eq:m-erlang}) 
is an increasing function of some linear combination 
of  $l(y)$  and $f(y)$ (at least for the examples of CDMA and OFDMA
given above). Indeed, we will show that the study of $\E[l(0)]$ and
$\E[f(0)]$ can explain the aforementioned non-monotonicity of the
blocking probability $\E[b]$ with  respect to the standard deviation of the shadowing.

\section{Numerical results}

\label{s.Numerical}
Following the methodology described in
Section~\ref{ss.methodologh}, in this section we will first
study the blocking probability and then the mean path-loss and
interference factors for the hexagonal and Poisson
network models with log-normal shadowing.

\subsection{Blocking probability}
\label{ss.NResBlocking}

In this section we consider only the hexagonal network on the torus
$\bbT_6$. We evaluate the blocking probability $\E[b]$ 
in  OFDMA network using the
Kaufman-Roberts algorithm, as described in Section~\ref{ss.Blocking}.

We assume the following parameter values for  OFDMA:
System bandwidth $W=5$MHz.
BS are equipped with omnidirectional 
antennas having a gain $9$dBi and transmit with
the maximal power $43$dBm; 
thus $\tilde{P}=43+9=52$dBm when we account for the isotropic antenna gain. 
The common channel power is the fraction $\epsilon=0.12$ of~$\tilde{P}$.
The ambient noise power is assumed
$N=-103$dBm.

We assume perfect  intra-cell orthogonality, i.e.,
$\alpha=0$.  Moreover we are not considering any  
opportunistic scheduling over fading. This allows us to 
characterize the link performance (averaged over fading) 
via the Shanon's formula 
$r/W=\psi\left(  \xi\right)  =\log_{2}\left(
1+\xi\right)$, without specifying into how many 
sub-carriers the bandwidth $W=5$MHz is split. 
 
We assume a traffic demand of 46.2~Erlang per km${}^2$ %
consisting of streaming calls at the bit-rate $r=180$Kbits/s
(typical for videoconferencing)
that is served by the hexagonal network consisting of 36 BS (on the tours
$\mathbb{T}_6$) with the distance between adjacent BS 
$\Delta=1$km. %

\begin{figure}[t]
\centering{
\includegraphics[width=0.8\linewidth]{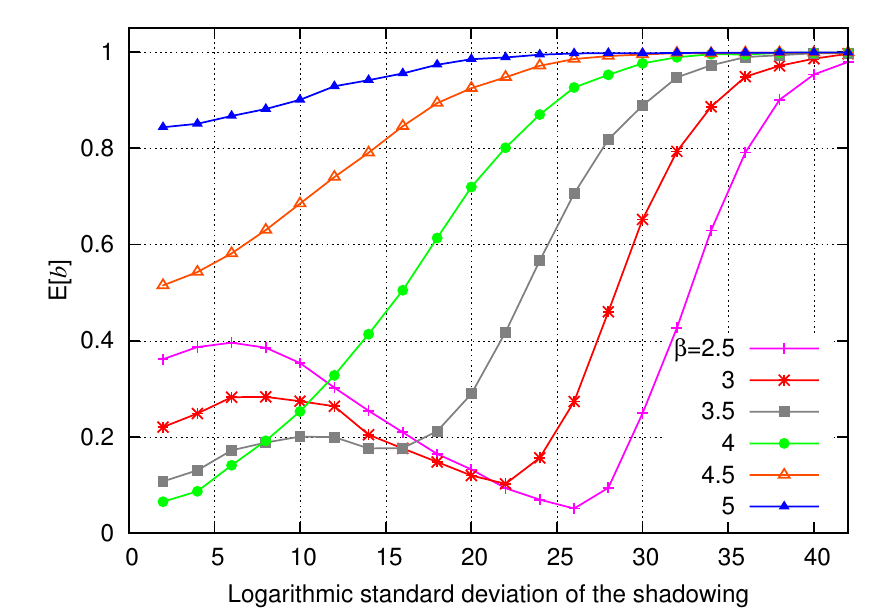}}
\caption[Blocking probability in OFDMA,  traffic 
46.2~Erlang per km${}^2$.]{\label{fig.BP-OFDMA-Hex}
Blocking probability in OFDMA
hexagonal network on the torus $\bbT_6$
with log-normal shadowing with 
log-SD $v$ and  path-loss exponent $\beta$, evaluated  
 using the Kaufman-Roberts algorithm for the traffic 
46.2~Erlang per km${}^2$.}
\end{figure}

The (deterministic) distance-loss function is  
$L(x)=\left(  Kx\right)  ^{\beta}$   with $K=8667$km${}^{-1}$
(which follows from Cost-Hata model~\cite{Cost231_1999}
for urban areas, assuming frequency 1795Mhz, BS antenna height 50m,
mobile antenna height 1.5m, for $\beta=3.38$).
Moreover, we assume that the values of 
the shadowing $S_X(y)$ for given $X$
and different locations $y$ are independent.
Figure~\ref{fig.BP-OFDMA-Hex}
shows the dependence of the blocking probability $\E[b]$ 
on the path-loss exponent~$\beta$ and the logarithmic standard
deviation $v$ of the log-normal shadowing evaluated in our  OFDMA network model.

\begin{Remark}\label{R.not-monoton}
\begin{itemize}
\item For negligible shadowing (logarithmic standard
deviation close to~0) the blocking probability $\E[b]$ first decreases in
the path-loss exponent~$\beta$ (on our figures
for $\beta$ from $2.5$ to~4) and
then increases in~$\beta$.
\item  The blocking probability is not always increasing in the
  standard deviation of the shadowing. Indeed, on our figures
with $\beta\ge 4 $ it is monotone increasing. However, 
for $\beta\le 3.5$
the blocking probability $\E[b]$ first increases, then decreases,
and ultimately increases to~1.
\end{itemize}
\end{Remark}
Note the decrease of the blocking probability in the standard
deviation of the shadowing can be quite
significant even between 7 and 15 dB,
depending on the path-loss exponent.

\begin{figure*}[t]
\centering{\ \hspace{-0.04\linewidth}\subfigure[OFDMA, 34.6~Erlang per km${}^2$]{\includegraphics[width=0.52\linewidth]{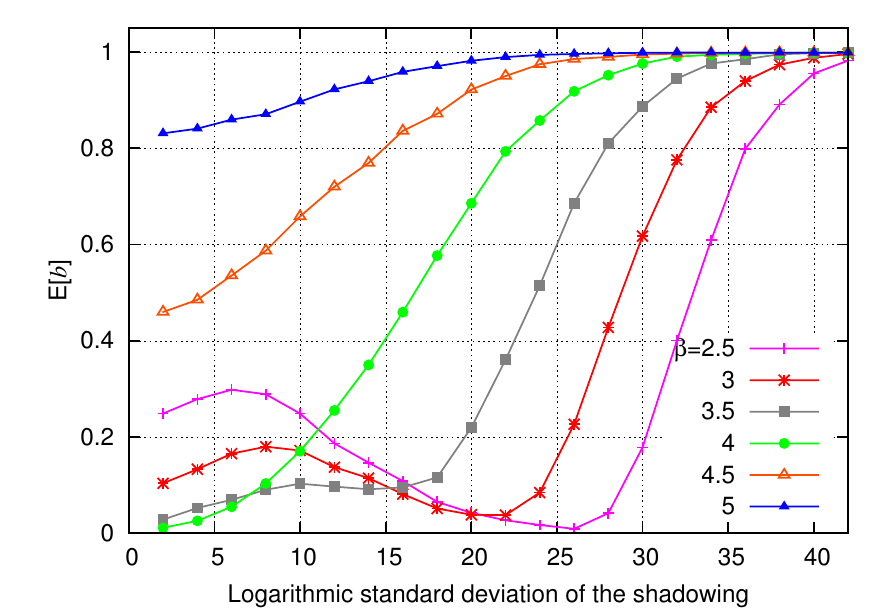}%
\label{fig.BP-OFDMA-hex-30}}\hfill\hspace{-0.5\linewidth}
\hfil
\subfigure[OFDMA, 23.1~Erlang  per km${}^2$]{\centering\includegraphics[width=0.52\linewidth]{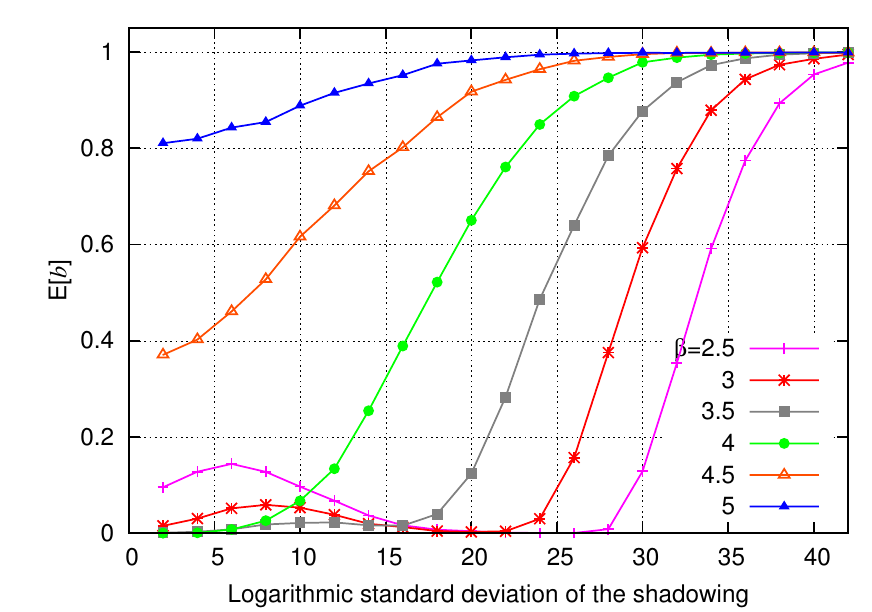}
\label{fig.BP-OFDMA-hex-20}}}
\caption[Blocking probability in OFDMA, traffic 
34.6 and 23.1~Erlang per km${}^2$.]{\label{fig.BP-OFDMA-hex-30-20}
Blocking probability for  OFDMA network as on
Figure~\ref{fig.BP-OFDMA-Hex} with traffic 
34.6~Erlang  per km${}^2$ and 23.1~Erlang per km${}^2$.
}
\end{figure*}

The lack of monotonicity observed in Remark~\ref{R.not-monoton}
is not specific for our choice of the traffic
of  46.2~Erlang par km${}^2$ as can be remarked on
Figure~\ref{fig.BP-OFDMA-hex-30-20}, where we have assumed two
different smaller values of the traffic. 
Moreover, we have observed very similar patterns, 
not presented here due to space constraints, in our model of CDMA.
Furthemore,  we have  confirmed these results by the
the crude Monte-Carlo simulations of the  network with the
arrivals and departures of users (implemented in MATLAB).

In the next section we will explain this behavior
and argue that it my be expected for other 
QoS metrics which depend on some combination of
the  path-loss factor and the interference factor.

\subsection{Analysis of the interference factor}
\label{ss.NResFFactor}

Now, we will study the impact of the
shadowing and also the geometry and size of the network on the
interference factor $\E[f(0)]$ that is a key to the understanding of the
strange non-monotonicity of the blocking probability shown above.
Recall that, contrarily to the blocking probability, the expectation
 $\E[f(0)]$ (as well as $\E[l(0)]$) does not depend
on any particular correlation  of the values of 
the shadowing $S_X(y)$ for given $X$
and different locations $y$.

We begin with an important observation made directly from our model.
\begin{Remark}\label{r.f-invariance}
By the homothetic invariance of our hexagonal and Poisson models
on the torus, or in the  infinite models, 
with the distance-loss function~(\ref{e.L}),  
the   {\em mean interference factor 
does not depend on the intensity~$\lambda$} of
BS but only on the size $N$ of the network.
\end{Remark}

\begin{figure*}[t]
\centering{
\ \hspace{-0.04\linewidth}\subfigure[hexagonal network]{\includegraphics[width=0.52\linewidth]{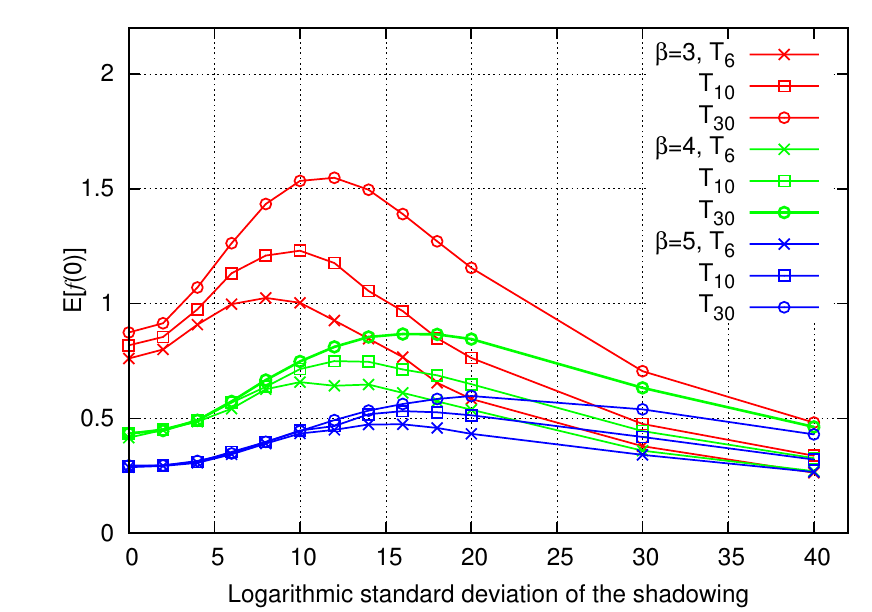}
\label{fig.fHex}}
\hfill\hspace{-0.05\linewidth}
\subfigure[Poisson network]{\centering\includegraphics[width=0.52\linewidth]{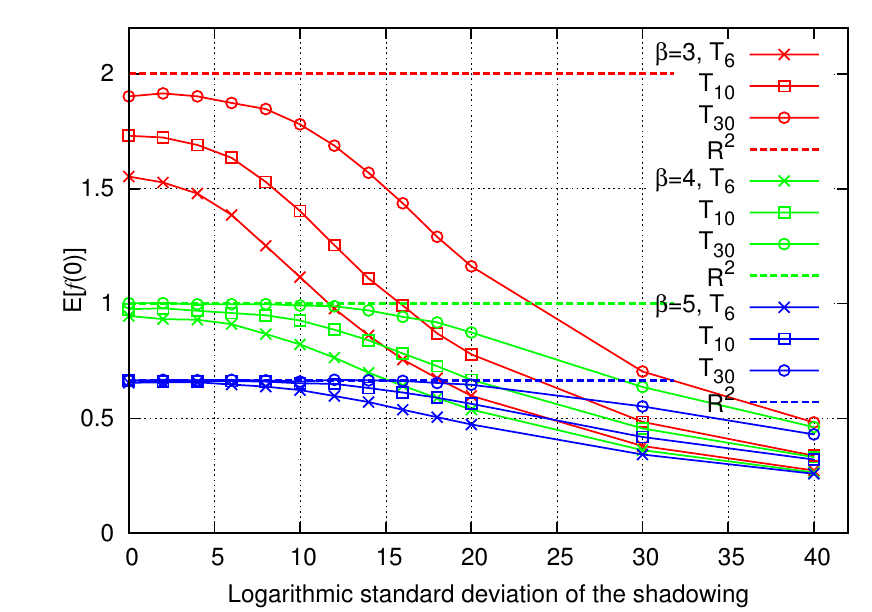}\label{fig.fPo}}}
\caption[Mean interference factor in hexagonal and Poisson
network.]{Mean interference factor in hexagonal and Poisson
network on the torus $\bbT_N$
with log-normal shadowing with 
log-SD $v$ and  path-loss exponent $\beta$.
Note that $\E[f(0)]$ increases with the size of the network.
The straight lines correspond to the infinite (on $\ir^2$) Poisson
model; cf. Proposition~\ref{p.poisson}.
}
\end{figure*}

Figures~\ref{fig.fHex} and~\ref{fig.fPo} show the impact  of the
path-loss exponent, shadowing
and the size of the network in the case of the hexagonal and Poisson
network architecture, respectively. Here are our main observations.
\begin{Remark}
\begin{enumerate}
\item Observe on Figure~\ref{fig.fHex}
{\em for hexagonal network} of a given size $N^2$~BS, with $N=6,10,30$, and
a given path-loss exponent~$\beta=3,4,5$, that {\em the mean interference
factor $\E[f(0)]$ first
  increases and then decreases to~0 when the  value $v$ of logarithmic
  standard deviation (log-SD) of the
  shadowing increases}.
\item {\em For the Poisson network} (see Figure~\ref{fig.fPo})
 $\E[f(0)]$ {\em decreases in log-SD 
  starting already from very small values of $v$}.
 \item The {\em actual size of the network consisting of $N^2$~BS, 
when $N\ge 100$, has
   negligible impact   on $\E[f(0)]$ when $\beta=4$ and $v\le 10$
or $\beta=5$  and $v\le 15$}  
both in hexagonal and Poisson case (in this latter case $N^2$ is the
expected number of BS).
In this regime the {\em value of $\E[f(0)]$ corresponds to this  in the
respective infinite model}. In particular, for Poisson network 
it is equal to $2/(\beta-2)$ and does not depend on log-SD $v$
(cf. Proposition~\ref{p.poisson} below).
\item When $\beta=4$ and $v\ge 10$
or $\beta=5$  and $v\ge 15$ the  
mean interference factor $\E[f(0)]$ non-negligibly 
increases with the  network size. 
\item Comparing Figures~\ref{fig.fHex} and~\ref{fig.fPo} for $v\ge 20$
  we observe that {\em for large log-SD of the shadowing
the mean interference factor evaluated for the
Poisson network is almost exactly  the same as for the hexagonal
network of the same size}.
\end{enumerate}
\end{Remark}

\begin{Remark} The seminal paper~\cite{ViterbiViterbiZehavi1994}
 considers only the hexagonal network architecture, however, the beneficial
 impact of the shadowing is not observed there. The reason is that 
the model considered in~\cite{ViterbiViterbiZehavi1994} 
 assumes that the smallest-path-loss BS
(the serving one) is selected among the $N_C$ closest BS.
In particular, $N_C=1$ ignores the shadowing in the handover policy
as it corresponds to the situation where the
serving BS is always the closest one. On the other hand the model
considered in our paper corresponds to  $N_C$ equal to the total
number of BS in 
the network. In consequence, for a higher path-loss exponent
(say $\beta=4$) and small and moderate log-SD of the shadowing ($0\le v\le
12$) our numerical results 
 are close to those of~\cite{ViterbiViterbiZehavi1994}
 with $N_C=4$; cf. our Figure~\ref{fig.fHex} and the last column in
 Table~1 in~\cite{ViterbiViterbiZehavi1994}. 
The fact that the average interference factor decreases in
some cases with log-SD of the shadowing has not been observed
in~\cite{ViterbiViterbiZehavi1994} due to the set of parameters considered
there. Indeed,  for a smaller path-loss exponent, $\beta=3$, 
our Figure~\ref{fig.fHex} shows the mean 
interference factor decreasing in $v$ starting from 
$v\approx 8$. This range of parameters is also considered
in~\cite[Table~2]{ViterbiViterbiZehavi1994}  however,
with the $N_C=2$. Apparently the beneficial impact of the shadowing
cannot be observed in this case, when the BS can be chosen only among
two closest BS. A general remark is of the following order: strong
shadowing requires larger geographical domain  in which the serving
BS is searched, as the optimal one may be located far from the mobile.
\end{Remark}

\subsection{Analysis of the path-loss factor}
\label{ss.NResLFactor}
\begin{figure*}[t]
\centering{
\ \hspace{-0.04\linewidth}\subfigure[hexagonal
network]{\centering\includegraphics[width=0.52\linewidth]{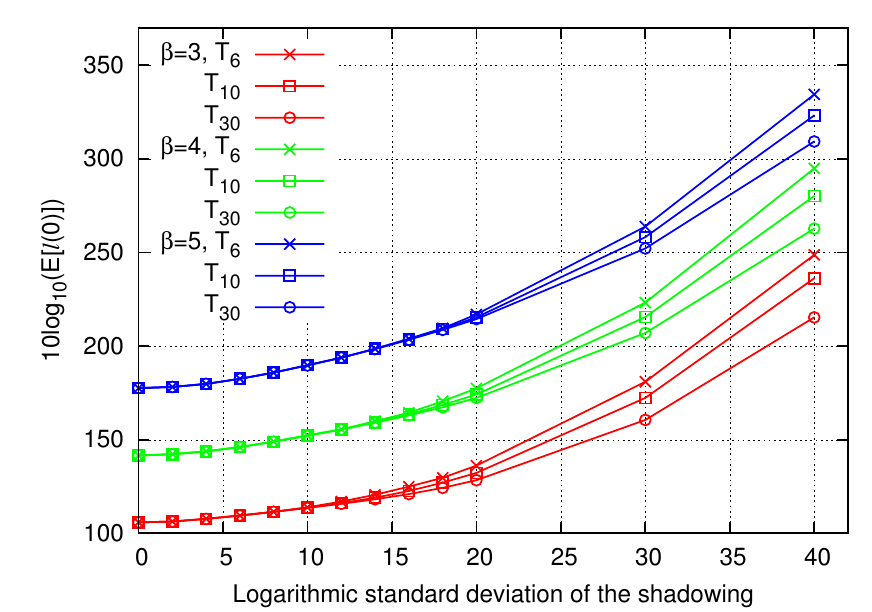}\label{fig.lHex}}
\hfill\hspace{-0.05\linewidth}
\subfigure[Poisson network]{\centering\includegraphics[width=0.52\linewidth]{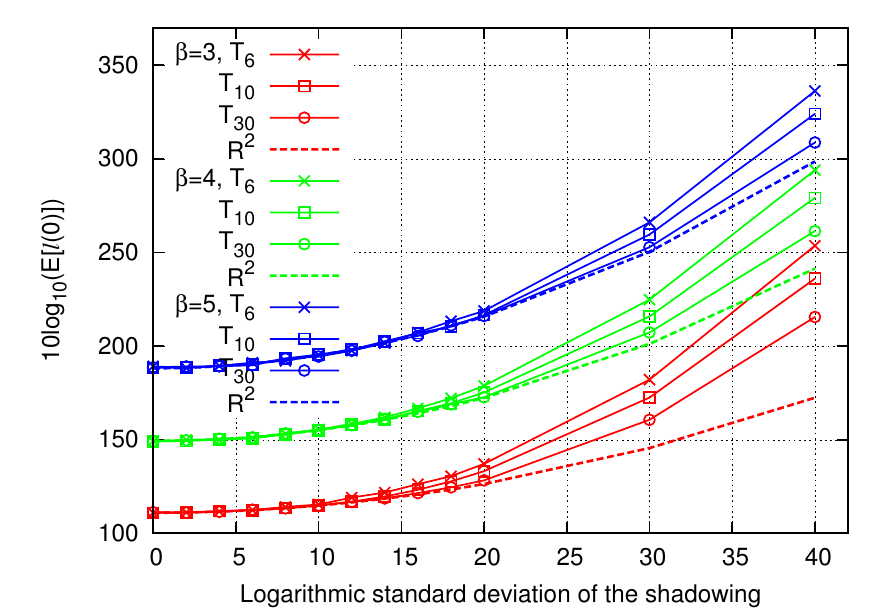}\label{fig.lPo}}}
\caption[Mean path-loss factor in hexagonal and Poisson network.]{
Mean path-loss factor  expressed in dB, in hexagonal and Poisson network 
on the torus $\bbT_N$ with log-normal shadowing with 
log-SD $v$ and  path-loss exponent $\beta$.
Note that $\E[l(0)]$ decreases with the size of the network.
The analytical expression for the  infinite (on $\ir^2$) Poisson
model is given in Proposition~\ref{p.poisson}.}
\end{figure*}

We begin with an important remark regarding the scaling of $\E[l(0)]$
with respect to the density of the BS.
\begin{Remark}\label{r.l-invariance}
Unlike the mean interference factor $\E[f(0)]$ (cf.
Remark~\ref{r.f-invariance}),
{\em the mean 
path-loss factor $\E[l(0)]$ depends on the intensity $\lambda$ 
of BS}. By the homothetic invariance of our hexagonal and Poisson
models, it is easy to see in the case of the  distance-loss
function~(\ref{e.L}) that this dependence has the following form 
$\E[l(0)]=\lambda^{-\beta/2}\Bigl(\E[l(0)]|_{\lambda=1}\Bigr)$.
Consequently, in particular, 
the path-loss factor becomes preponderant in the case of
sparse networks (small $\lambda$) and negligible for dense
networks (large $\lambda$). We will see in
Section~\ref{ss.ffactorPoisson} that  $\E[l(0)]$ can be evaluated
explicitly in the case of the infinite Poisson network with an arbitrary
distribution of the shadowing. 
\end{Remark}
Figures~\ref{fig.lHex} and~\ref{fig.lPo} show 
the mean path-loss factor $\E[l(0)]$
evaluated for the intensity of BS
$\lambda=1.155$BS/km${}^2$ (equivalent to $\Delta=1$km).
The main observations are presented in the next section.

\subsection{Conclusions on numerical results}
\begin{figure}[t]
\centering\includegraphics[width=0.5\linewidth]{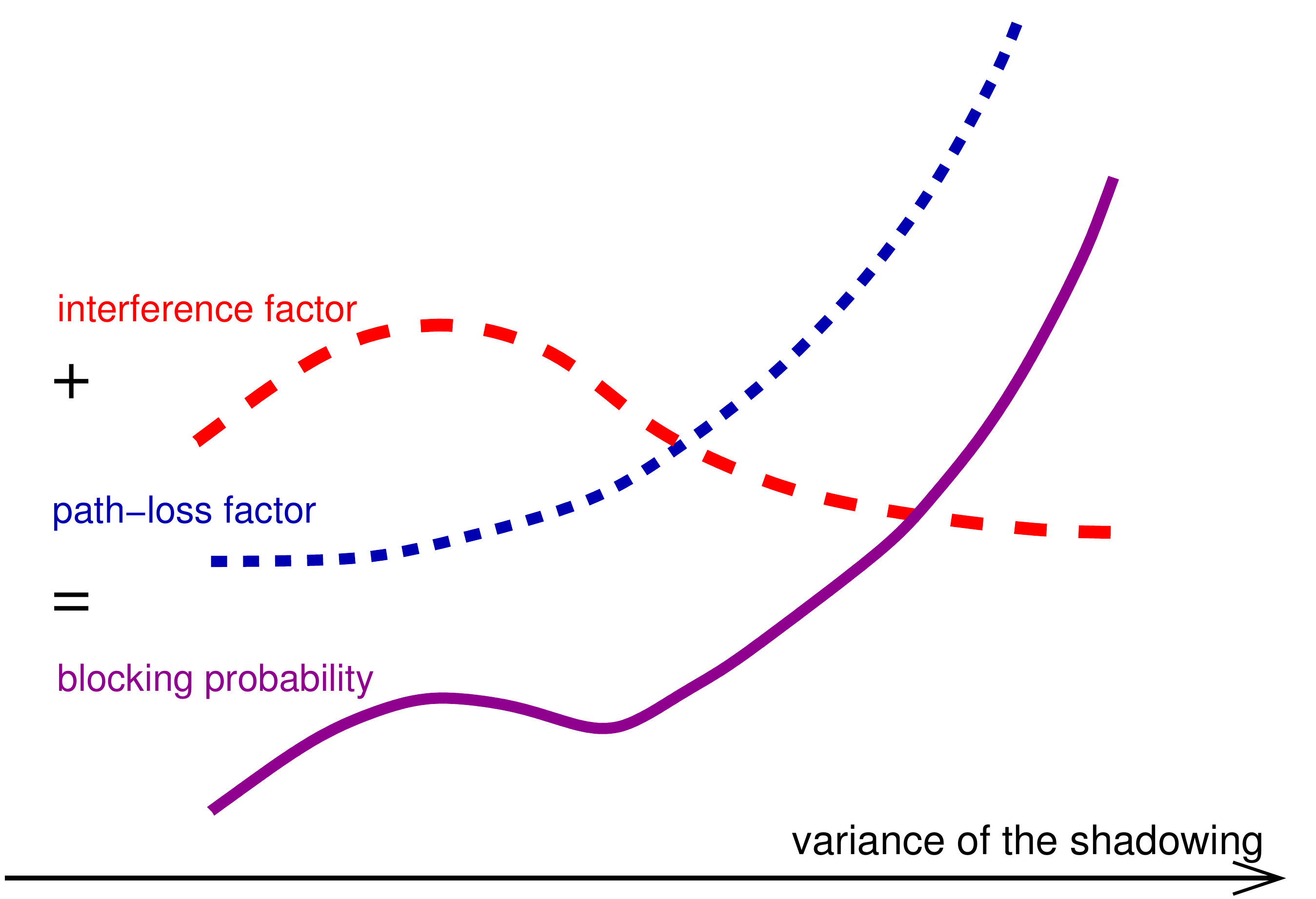}
\caption[Blocking probability versus variance of the shadowing.]{\label{fig.tendency}
Graphical explanation 
of a possible shape of the dependence of 
the blocking probability on  the variance of the shadowing.}
\end{figure}

For the hexagonal network we have observed the following 
facts regarding our two QoS ``pre-metrics''.
\begin{itemize}
\item   The mean path-loss factor increases to infinity 
in  the standard deviation of the shadowing,  increases in the 
path-loss exponent, increases in the cell radius, 
but (slightly)  decreases in the number of base stations.
\item The mean interference factor is not monotone in
the   standard deviation of the shadowing: first
  increases and then decreases to~0.
It decreases in the path-loss
 exponent, is invariant with respect to the cell radius
and  increases in the number of base stations.
\end{itemize}
Knowing that the blocking probability  depends
on some combination of the path-loss and interference factors of
users (cf. formulas~(\ref{eq.CDMA-phi}) and~(\ref{eq.OFDMA-phi})), 
and having observed that the mean values of these two factors
have opposite monotonicities in the path-loss exponent, 
it is not surprising that the blocking probability is not monotone in
the path-loss exponent; cf. the first observation of Remark~\ref{R.not-monoton}.
A similar argument explains a possible non-monotonicity of the
blocking probability in the standard deviation of the shadowing;
cf. the second  observation of Remark~\ref{R.not-monoton} and the
scheme on Figure~\ref{fig.tendency}.

For the Poisson network we have observed the same tendencies
of QoS ``pre-metrics'' as for hexagonal network mentioned above, except that  
the mean interference factor is monotone decreasing in the
shadowing. Moreover,  for large standard deviation of the shadowing,
the  ``pre-metrics'' of the Poisson network are
very close to those of the hexagonal network. 

In the next section we will prove also that 
for the infinite Poisson  network 
the distributions of our QoS ``pre-metrics'' do not depend on the
shadowing and admit explicit formulas for their means.

\section{Mathematical results}
\label{s.Mathematical}
In this section we will state and  prove some
mathematical results regarding $\E[l(0)]$ and 
$\E[f(0)]$, which support and extend
the  numerical findings of Section~\ref{s.Numerical}.

\subsection{Toroidal models}

We begin by a simple observation regrading the log-normal distribution
 of the shadowing $S$ with mean~1. Recall, it can be represented as
 $S=e^{-\sigma^2/2+\sigma N}$ where $N$ is the standard Gaussian random variable.
Thus,  for any fixed $\epsilon>0$ we have 
\begin{eqnarray*}
\Pr\{S\ge \epsilon\}&=&\Pr\{N\ge \sigma/2+(\log\epsilon)/\sigma\}
\buildrel{\sigma\to\infty}\over\longrightarrow 0\,,
\end{eqnarray*}
which shows that the random variable $S$ {\em converges in probability
  to $0$} 
when $\sigma$ (and hence $v=\sigma 10/\log10$) tends to infinity
(and this even if $\E[S]\equiv 1$!). From this,  we have that the {\em
  path-loss
between any location $y$ and any  BS $X$, $L_X(y)=
L(|X-y|)/S_{X}(y)$, converges in probability to infinity when 
the variance of the shadowing increases}. Consequently,
for any {\em finite} network $\tilde\Phi$ of base stations,
the {\em path-loss factor $l(y)=\min_{X\in\Phi}L_X(y)$ converges in
  probability and in expectation to infinity}. 
This explains the asymptotics  of $\E[l(0)]$ for large $v$
observed on Figures~\ref{fig.lHex} and~\ref{fig.lPo}.

The somewhat surprising observation on Figures~\ref{fig.fHex} and~\ref{fig.fPo}
regarding the
beneficial impact of the strong log-SD $v$ of the shadowing on the mean
interference factor can be also confirmed mathematically.
\begin{Prop}\label{p.fto0}
Assume an arbitrary,  fixed, finite  pattern $\{X_1,X_2,\ldots,X_n\}$
of BS locations. Consider any deterministic distance-loss function
$0<L(r)<\infty$ and (independent) log-normal shadowing
$S_{X_i}(\cdot)$  with the log-SD $v$. 
Then for any location $y$ we have
$\lim_{v\to\infty} f(y)=0$ in probability.
\end{Prop}  
\begin{proof}
It is enough to show  $\lim_{v\to\infty}\Pr\{\,f(y)\ge \epsilon\,\}=0$
for any $\epsilon$ satisfying $0<\epsilon<1$.
Denote by $G_i=S_{X_i}(y)/L(|X_i-y|)$ the path-gain from $X_i$ to $y$.
Consider ordered vector  $(G_{(1)}, \ldots,G_{(n)})$ of these path
gains, where $\min_iG_i=G_{(1)}\le\ldots\le G_{(n)}=\max_{i}G_i$.   
Note that $f(y)=1/G_{(n)}\sum_{i=1}^nG_{(i)}-1\le
(n-1)G_{(n-1)}/G_{(n)}$. In order to prove our claim it is enough to show
that $\Pr\{\,G_{(n-1)}/G_{(n)}\ge \epsilon\,\}\to0$ when $v\to\infty$.
To this regard denote $L(|X_i-y|)=l_i$, and
recall from the definition of  our path-loss model 
that we can represent  $G_i(y)=e^{\tilde N_i}$, where 
$\{\tilde N_i\}_{i=1,\ldots,n}$ are independent Gaussian random
variables, with mean $\E[\tilde N_i]=-\log l_i-\sigma^2/2$ and the same
SD $\sigma=v\log10/10$. Since $G_i$ is monotone increasing in $\tilde
N_i$ we have $G_{(i)}=e^{\tilde N_{(i)}}$,   where $\min_i\tilde
N_i=\tilde N_{(1)}\le\ldots\le \tilde N_{(n)}=\max_{i}\tilde N_i$.
Moreover,  
$A:=\{\,G_{(n-1)}/G_{(n)}\ge \epsilon\,\}=
\{\,\tilde N_{(n)}-\tilde N_{(n-1)}\le M\,\}$,
where $M=- \log \epsilon$.
Denote by $A_{ij}=\{\,0\le \tilde N_i-\tilde N_j\le M\,\}$.
Note that $A\subset\bigcup_{i,j=1,\ldots,n, i\not=j}A_{ij}$
and the result follows from the fact that for any $i\not=j$
$\Pr\{A_{ij}\}\to0$ when $v\to\infty$.
Indeed, for $i\not=j$, $\tilde N_i-\tilde N_j=\bar N$ is Gaussian random
variable with mean $\log (l_j/l_i)$ and variance $\sigma^2$
and thus $\Pr\{A_{ij}\}=\Pr\{\,0\le\bar N\le M\}\to0$ for any given
finite $M$ when $\sigma^2=v^2\log^210/100\to\infty$.
This completes the proof.
\end{proof}
\begin{Cor}\label{c.Efto0}
Assume Poisson or hexagonal network on the torus $\bbT_N$, with
log-normal shadowing having log-SD~$v$. Then
the mean interference factor $f(0)$  
converges in distribution and in expectation 
to 0 when $v\to\infty$.
\end{Cor}
\begin{proof}
For any $\epsilon>0$, by Proposition~\ref{p.fto0} and Lebesgue dominated
convergence theorem we have $\Pr\{\,f(0,\tilde\Phi)>\epsilon\,\}
=\E[\Pr\{\,f(0,\tilde\Phi)>\epsilon\,|\Phi\}]\to0$, when $v\to\infty$.
This proves that $f(0)$ converges in distribution to~0. Convergence of
$\E[f(0)]$ to 0 follows again from the Lebesgue dominated
convergence theorem by the observation
$f(y,\tilde\Phi)\le\Phi(\bbT_N)-1$ and $\E[\Phi(\bbT_N)]<\infty$.
\end{proof}

\begin{Remark}\label{R.big-jump}
Recall that the log-normal distribution of the shadowing is heavy-tailed.
The result of Proposition~\ref{p.fto0} and Corollary~\ref{c.Efto0} 
can be heuristically explained and conjectured for other heavy tailed
shadowing distributions by the so called  {\em single big jump
  principle}, cf e.g.~\cite[Section~3.1]{Foss-etal2011}. It says that the
only significant 
way in which a large value of the sum of independent 
heavy-tailed variables can be attained is
through a big value of single  term of the sum  (``big jump'').
In other words, the maximum and the sum of independent 
heavy-tailed random variables have the same asymptotic
of the distribution function for large values. Note also, that we
observe the ``single big jump principle'' 
in a different scenario: we study the ratio of the interference
(sum of the log-normal 
path-gains minus the largest path-gain) to the serving-BS path-gain
(the largest one) {\em asymptotically for large variance}.
\end{Remark}

\subsection{Infinite models}
\label{ss.ffactorPoisson}

In this section we will consider infinite hexagonal and Poisson 
models. We will show first that serving BS $X_0^*$, and hence the path-loss
and interference factors, are well defined. 
Then we will argue that values of these factors
in the infinite models  can be seen as limits
of respective  toroidal models  on $\bbT_N$ when $n\to\infty$.
Finally we will prove a (surprising?) invariance of $\E[l(0)]$ and
 $\E[f(0)]$ in  the infinite Poisson model  
with respect to the distribution of the shadowing.
In this case the values $\E[l(0)]$ and $\E[f(0)]$ can
be evaluated explicitly.

\begin{Prop}\label{p.welldefined}
Consider infinite Poisson $\Phi=\Phi_P$ or hexagonal 
$\Phi=\Phi_H$ model of BS, with shadowing
whose marginal distribution has finite moment of order $2/\beta$
(\footnote{i.e., $\E[S^{2/\beta}]<\infty$. Note that $2/\beta<1$ and
  thus the above assumption follows from  our default  assumption
  $\E[S]=1<\infty$.}).
Then there exist $X_0^*\in\Phi$ satisfying~(\ref{e.handover}). 
Moreover, the path-loss factor and the 
interference factor calculated with respect to 
the restriction of $\Phi$ to $\bbT_N$, i.e.,  $l(0,\tilde\Phi^{\bbT_N})$
and $f(0,\tilde\Phi^{\bbT_N})$, 
converge almost surely and in expectation to $l(0,\tilde\Phi)$
and $f(0,\tilde\Phi)$, respectively.
\end{Prop}
\begin{proof}
To prove the first statement it is enough to show that
the expected number of BS $X_i$ such that $S_{X_i}(0)/L(|X_i|)> M$
is finite for any $M<\infty$. In the case of the Poisson p.p. this
will be shown in the proof of  Proposition~\ref{p.poisson} below.
Here we consider only hexagonal case $\Phi=\Phi_H$.
Denote by $\overline G(x)=\Pr\{\,S> x\,\}$. We have 
\begin{eqnarray*}
\lefteqn{\E[\#\{X_i\in\Phi_H:S_{X_i(0)}/L(|X_i|)> M\}]}\\
&=&
\E\Bigl[\sum_{X_i\in\Phi_H}\ind\Bigl(S_{X_i(0)}>ML(|X_i|)\Bigr)\Bigr]\\
&=&\E\Bigl[\sum_{X_i\in\Phi_H}\overline G\Bigl(ML(|X_i|)\Bigr)\Bigr]\\
&\le&\sum_{i=1}^\infty 6n
\overline G\Bigl((n\Delta
K/2)^\beta/M\Bigr)<\infty,
\end{eqnarray*}
\label{ind.f-page}
where $\ind(\cdot)$ denotes the indicator function and 
the last inequality follows from the assumption  
$\E[S^{2/\beta}]=2/\beta\int_0^\infty s^{2/\beta-1}\overline G(s)\,ds<\infty$.
This completes the proof of the first statement. 

In order to prove the second statement, note that for any realization
the network $\tilde\Phi$, for $N$ large enough
$X_0^*\in \bbT_N$. Consequently, $l(0,\tilde\Phi^{\bbT_N})$
is eventually constant in $N$ while 
 $f(0,\tilde\Phi^{\bbT_N})$
eventually increases in $N$ (the serving BS is not changing any more and only
interference is added). The convergence of the expectation of the path-loss factor
follows from the monotone convergence theorem, noting that 
$l(0,\tilde\Phi^{\bbT_N})$ is decreasing in $N$.
The convergence of the expectation of the interference  factor follows
form the    
dominated convergence theorem
 knowing that 
$f(0,\tilde\Phi)\le f'(0,\tilde\Phi)$, where $f'(0,\tilde\Phi)$ is 
the interference factor calculated under assumption that  the handover
policy selects the geographically closest BS  as the serving one.
By the independence of the shadowing fields given the locations
of BS and the assumption that the mean shadowing is equal to~1 
\begin{equation}\label{e.fprim}
\E[f'(0,\tilde\Phi)]=\E\Bigl[\frac{1}{S}\Bigr]\E\Bigl[\sum_{X\in
  \Phi}\frac{L(|X_0^{'*}|)}{L(|X|)}\Bigr]-1\,,
\end{equation}
where $X_0^{'*}$ is a point of $\Phi$ closest to the origin~0.
By our assumption on the mean path-gain $\E[1/S]<\infty$.
The second expectation~(\ref{e.fprim}) is
equal to the mean  interference factor in the infinite model
with constant shadowing $S\equiv1$, and it is known to be finite in
the infinite hexagonal and Poisson model;
cf. respectively Remark~\ref{r.HexInfini} and
Proposition~\ref{p.poisson} below. 
\end{proof}

\begin{Remark}\label{r.HexInfini}
It was shown
in~\cite{BaccelliBlaszczyszynKarray2004} that
in the case of $S\equiv1$ and the deterministic distance-loss
function~(\ref{e.L}) $\E[l(0)]$ and $\E[f(0)]$ in the hexagonal model
can be  approximated by  the following expressions 
\begin{eqnarray*}
\E[f(0,\Phi_H)]&\approx &\frac{0.9365}{\beta-2},\\
\E[l(0,\Phi_H)]&\approx &\frac{K^\beta}{(\pi\lambda)^{\beta/2}(1+\beta/2)}\,.
\end{eqnarray*}
\end{Remark}
To the best of our knowledge, 
analytical expressions (approximations) in the case of the infinite
 hexagonal network with random
shadowing are not known.
We consider now infinite Poisson model.
\begin{Prop}\label{p.poisson}
Assume infinite Poisson network, deterministic distance-loss function~(\ref{e.L})
and a general distribution of the shadowing $S$ satisfying 
$\E[S^{2/\beta}]<\infty$. Then 
the distribution of the interference factor $f(0)=f(0,\tilde\Phi)$
does not depend on the distribution 
$S$ and the distribution of $l(0)$ depends on $S$ only through the
product  $\lambda\E[S^{2/\beta}]$.
Moreover
\begin{eqnarray*}
\E[f(0)]&=&\frac{2}{\beta-2}\,,\\
\E[l(0)]&=&\frac{K^\beta 
\Gamma(1+\beta/2)}{(\pi\lambda\E[S^{2/\beta}])^{\beta/2}}\,,
\end{eqnarray*}
where $\Gamma(a)=\int_0^\infty t^{a-1}e^{-t}dt$.
In particular, for the log-normal shadowing
$$\E[l(0)]=\frac{K^\beta\Gamma(1+\beta/2)\exp[(1-2/\beta)\sigma^2/2]}%
{(\pi\lambda)^{\beta/2}}\,.$$
\end{Prop} 
\begin{Remark}
The above result says that in the infinite Poisson notwork the
existence of shadowing  has no impact on 
the mean interference factor.
The impact of the shadowing on the mean path-loss factor in
this model consists in a ``fictitious'' scaling of the intensity of the BS
by the factor $(\E[S^{2/\beta}])^{\beta/2}\le1$.
The respective expressions in the case of $S\equiv 1$
has been found for the first time 
(to the best of our knowledge)
in~\cite{BaccelliBlaszczyszynTournois2003}. 
Note however, that the above observation is valid only if the handover
policy selects the  BS with the smallest path-loss, as described in
Section~\ref{ss.handover}.
Indeed, assume that, despite non-constant shadowing,  the handover
policy selects the geographically closest BS as the serving one.
Then, the mean interference factor $\E[f'(0)]$ can be expressed as
in~(\ref{e.fprim}). Recall that the second expectation in this
expression is equal to the mean interference factor in the same model
without shadowing (i.e., $S\equiv1$). By the Jensen's inequality
$\E[1/S]\ge 1/\E[S]=1$ and consequently we observe 
the increase of the mean interference factor compared to the
``shadowing-dependent'' handover policy.
In particular, for log-normal $S$ with mean~1 and log-SD ~$v$
we have $\E[1/S]=e^{\sigma^2}=e^{v^2\log^210/100}$, which means that the
{\em log-normal shadowing in any geometric model of BS in which it is not
taken into account in the handover policy increases the 
mean interference factor by}  $v^2\log10/10\,$dB, 
where $v$ is log-SD of the shadowing.
\end{Remark}

\begin{proof}[Proof of Proposition~\ref{p.poisson}]
Note that the values of $l(0)$ and $f(0)$ are
 entirely defined by the collection of
random variables $\{L_X(0)=L(|X|)/S_{X}(0): X\in\Phi\}$. Given $\Phi$ 
these random variables are independent. Thus by the displacement
theorem for Poisson
p.p. (cf.~\cite[Theorem~1.3.9]{FnT1})
$\{L_X(0)\}=\Psi$  constitutes a (non-homogeneous) Poisson p.p. on
$\ir^+=[0,\infty)$ of intensity measure $\Lambda'$ given by 
\begin{eqnarray*}
 \Lambda'([0,s])&=&\E[\Psi([0,s])]\\
&=&\lambda\int_{\ir^{2}}\Pr\{\,L(|z|)/S\le s\,\}\,dz\\
&=& 2 \pi \lambda \int_{0}^{\infty}r\Pr\{\,L(r)/S \le s \,\}\,dr\\
&=& 2 \pi \lambda \int_{0}^{\infty}r\E\left[\ind\left(L(r)/S\le s \right)\right]\,dr\\
&=& 2 \pi \lambda \E \left[ \int_{0}^{(sS)^{1/\beta}/K}r\,dr \right]\\
&=& \frac{ \lambda s^{2/\beta}\pi}{K^2} \E \left[ S^{\frac{2}{\beta}} \right]\,.
\end{eqnarray*}
Note that the latter expression is finite, which 
proves that the serving BS $X_0^*$ is well defined
(cf. proof of Proposition~\ref{p.welldefined}).
Note also that it depends on the shadowing only through 
its moment $\E[S^{2/\beta}]$. Moreover one obtains the same expression 
in the model without shadowing  and the density of
BS multiplied  by $\E[S^{2/\beta}]$.
By the homothetic invariance of the Poisson model
with the distance-loss function~(\ref{e.L}) the distribution of $f(0)$ does
not depend on the intensity of the BS. Thus the  invariance of the
distribution of $f(0)$ on the distribution of the shadowing. In
particular, we can conclude that $\E[f(0)]=2/(\beta-2)$ --- the value
obtained in the model without shadowing;
see~\cite{BaccelliBlaszczyszynTournois2003}, 
cf. also \cite[Example~4.5.1]{FnT1}. 
The formula for the mean path-loss factor follows from its dependence
on the intensity of the base stations via the function 
$\lambda^{-\beta/2}$.
This completes the proof.
\end{proof}

\section{Concluding remarks}
\label{s.Concluding}
We show that the QoS in path-loss-and-interference limited cellular 
networks is not always decreasing in the strength (variance) of the log-normal
shadowing, provided  
the handover policy selects the  BS with the smallest path-loss 
as the serving one.
Under strong shadowing it principally suffers from the
poor path-loss conditions with respect to the serving BS.
For moderate shadowing however, when the QoS is  not yet compromised by
the path-loss conditions, it may profit from
the reduction of the interference. This is because increasing variance of the
log-normal shadowing tends to ``separate'' 
the strongest (serving BS) signal from all other signals ---
the phenomenon observed for  heavy-tailed distributions 
and called  ``single big jump principle''. 
This mathematical result seems also to be in line with a recent 
real-network observation~\cite{NadineMALHOUROUX}
that mobiles in indoor communications  (typically subject to strong
shadowing) report fewer BSs for potential handover.
The results presented in this paper
regard the network-average of the QoS metrics.
More study is needed, to analyze the   impact of the shadowing on the 
distribution of these metrics in the network. This requires
appropriate models of the spatial correlation of the shadowing.

\addcontentsline{toc}{section}{References}
%\bibliographystyle{IEEEtranTCOM}
%\bibliography{Shadowing_Jnl}
% Generated by IEEEtranTCOM.bst, version: 1.13 (2008/09/30)

\end{document}